\documentclass{moriond}

\def\be{\begin{equation}}
\def\ee{\end{equation}}
\def\bea{\begin{eqnarray}}
\def\eea{\end{eqnarray}}

\usepackage{arydshln} 
\graphicspath{{figs/}}

\newcommand{\Photo}{}

\begin{document}

\begin{flushright}
 LPT-Orsay-18-65
\end{flushright}

\vspace*{.3cm}
\title{Heavy Neutral Leptons and displaced vertices at LHC}

\author{\sc Xabier Marcano }

\address{Laboratoire de Physique Th\'eorique, CNRS, \\
Univ. Paris-Sud, Universit\'e Paris-Saclay, 91405 Orsay, France}

\maketitle
\abstracts{
Heavy neutral leptons  are present in many well-motivated beyond the Standard Model theories, sometimes being accessible at present colliders. 
Depending on their masses and couplings they could be long-lived and lead to events with displaced vertices, and thus to promising signatures due to low Standard Model background.
We revisit the potential of the LHC to discover this kind of new particles via searches for displaced vertices, which can probe masses of few GeV for mixings currently allowed by experimental constraints.
We also discuss the importance of considering all the possible production channels, including the production in association with light neutrinos, in order to fully explore this region of the parameter space.
}

\section{Introduction}

Beyond the Standard Model (BSM) theories have the difficult challenge of solving the problems that the Standard Model (SM) cannot explain. 
Many well-motivated BSM theories, such as the type-I seesaw model or its variants, introduce heavy neutral leptons (HNL) to the particle spectrum in order to solve open problems  as the origin of neutrino masses, to provide a dark matter candidate or to generate the observed amount of baryon asymmetry  via leptogenesis. 

Nevertheless,  there is not a clear theoretical input for the mass scales of these new particles. 
Actually, in the case of models with more than one HNL, each of them introduces, in principle, a new mass scale. 
Consequently, one needs to explore the full mass range, both with cosmological observations and in the lab, and study the different phenomenology associated to each mass regime~\cite{Atre:2009rg,Abada:2017jjx}. 
For instance, if they are very light, they could affect neutrino oscillations, as suggested by some neutrino oscillation anomalies. 
In contrast, if they are very heavy, experiments at the high intensity frontier could probe their low-energy imprints, such as lepton flavor violating processes. 

The HNL could also be directly produced in different experiments, depending again on their masses. 
If they are light enough, they could be produced in nuclear $\beta$ decays or in meson decays, which could lead to experimental signatures as monochromatic lines or kinks in the associated electron or muon spectrums.
For heavier masses, they could be produced at higher energy colliders as the LHC, able  to probe masses in the range from few GeV  up to the TeV scale. 

It is interesting to have a closer look to this lower LHC mass regime of few GeV since, given current experimental constraints, it corresponds to the region of parameter space where the HNL could be long-lived and travel a macroscopical distance before decaying inside the LHC detectors. 
In such situation, the HNL could not be seen in generic prompt decay searches, and dedicated analyses looking for displaced vertices (DV) are needed. 
Moreover, this kind of DV signatures are mainly SM background free, meaning that the observation of few events would be enough for a discovery.

In the last years, several studies have pushed for this kind of dedicated LHC searches for HNL with charged leptons\,\cite{Nemevsek:2011hz,Helo:2013esa,Izaguirre:2015pga,Dube:2017jgo,Cottin:2018kmq,Cottin:2018nms,Dib:2018iyr,Nemevsek:2018bbt}, from Higgs decays\,\cite{Maiezza:2015lza,Gago:2015vma,Accomando:2016rpc,Nemevsek:2016enw,Caputo:2017pit,Deppisch:2018eth}, at the LHCb\,\cite{Antusch:2017hhu} or at future LHC detectors proposed for searching for long-lived particles\,\cite{Caputo:2016ojx,Kling:2018wct,Helo:2018qej,Jana:2018rdf,Curtin:2018mvb}.
Our aim\,\cite{Abada:2018sfh} is to join these efforts by revisiting HNL searches via DV signatures at the LHC. 
We explore the different neutrino production mechanisms, considering not only $W$ mediated channels, but also the ones via $Z$ or $H$ bosons, and compare among the discovery potentials for each of them. 
As we will see, the conclusion depends on the flavor structure of the HNL, implying that all these channels are complementary and necessary to probe the whole parameter space (masses of the HNL and their mixings to the SM active neutrinos).

\section{The 3+1 model}

As already mentioned, there are many BSM theories that add HNL to the SM particle content.
Depending on the kind of open problem they want to address, they introduce a different number of HNL or a particular mixing pattern to the SM neutrinos. 
In order to be as generic as possible, in this work we follow a bottom-up approach rather than choosing a specific model. 

We consider the SM with three massive light neutrinos as required by  oscillation phenomena, and add one sterile neutrino to its spectrum. 
We do not impose that the new sterile state is the responsible of generating light neutrino masses, which would be model dependent,   we assume instead that its mass $m_N$ and mixings with the light neutrinos, $V_{\ell N}$ with $\ell=e,\mu,\tau$, are free independent parameters. 

The only condition we impose is that the full $4\times4$ lepton mixing matrix in this $3+1$ model,
\begin{equation}
U_\nu^{3+1} = \left(\begin{array}{ccc:c}
&&& V_{eN} \\[.5ex]
\multicolumn{3}{c:}{\tilde U_{\rm PMNS}} & V_{\mu N}\\[.5ex]
&&& V_{\tau N}\\[.5ex]
\hdashline
V_{N e} & V_{N \mu} & V_{N \tau} & V_{NN}
\end{array}
\right)\,,
\end{equation}
is unitary. Here the $3\times 3$ $\tilde U_{\rm PMNS}$ matrix is similar to the usual PMNS matrix up to small non-unitarity corrections due to the presence of light-sterile $V_{\ell N}$ mixings.
These three mixings define the interaction strength of the HNL via charged currents, as well as the neutral currents to both $Z$ and $H$ bosons.
Therefore, they will be, together with the $m_N$ mass, the relevant parameters for our study.

\section{HNL production and decay at the LHC}

In order to have DV signatures from the HNL, its decay length needs to be of the same order of the size of the detector.
In the case of detectors such as ATLAS or CMS, DV could be seen for displacements between roughly 1~mm and 1~m from the interaction point.

When the HNL is very massive, $m_N>m_W$, it tends to decay very fast and, thus, we need to search for prompt decayed signatures.
On the other hand, if it is lighter, its decay happens mainly via off-shell $W$ or $Z$ bosons, leading to leptonic or semileptonic channels~\cite{Atre:2009rg,Abada:2017jjx,Bondarenko:2018ptm}.

\begin{figure}[t!]
\begin{center}
\includegraphics[width=0.49\textwidth]{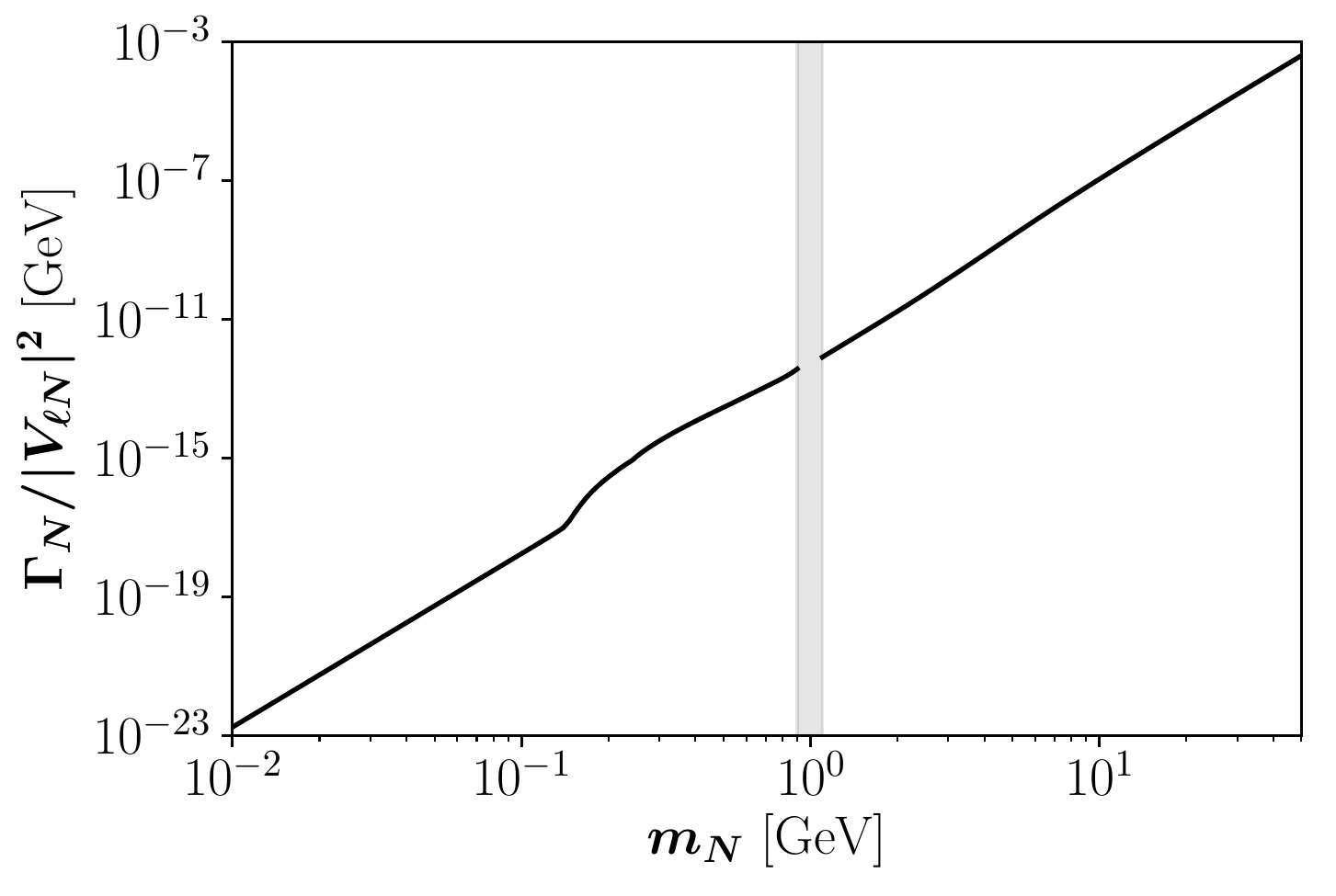}
\includegraphics[width=0.49\textwidth]{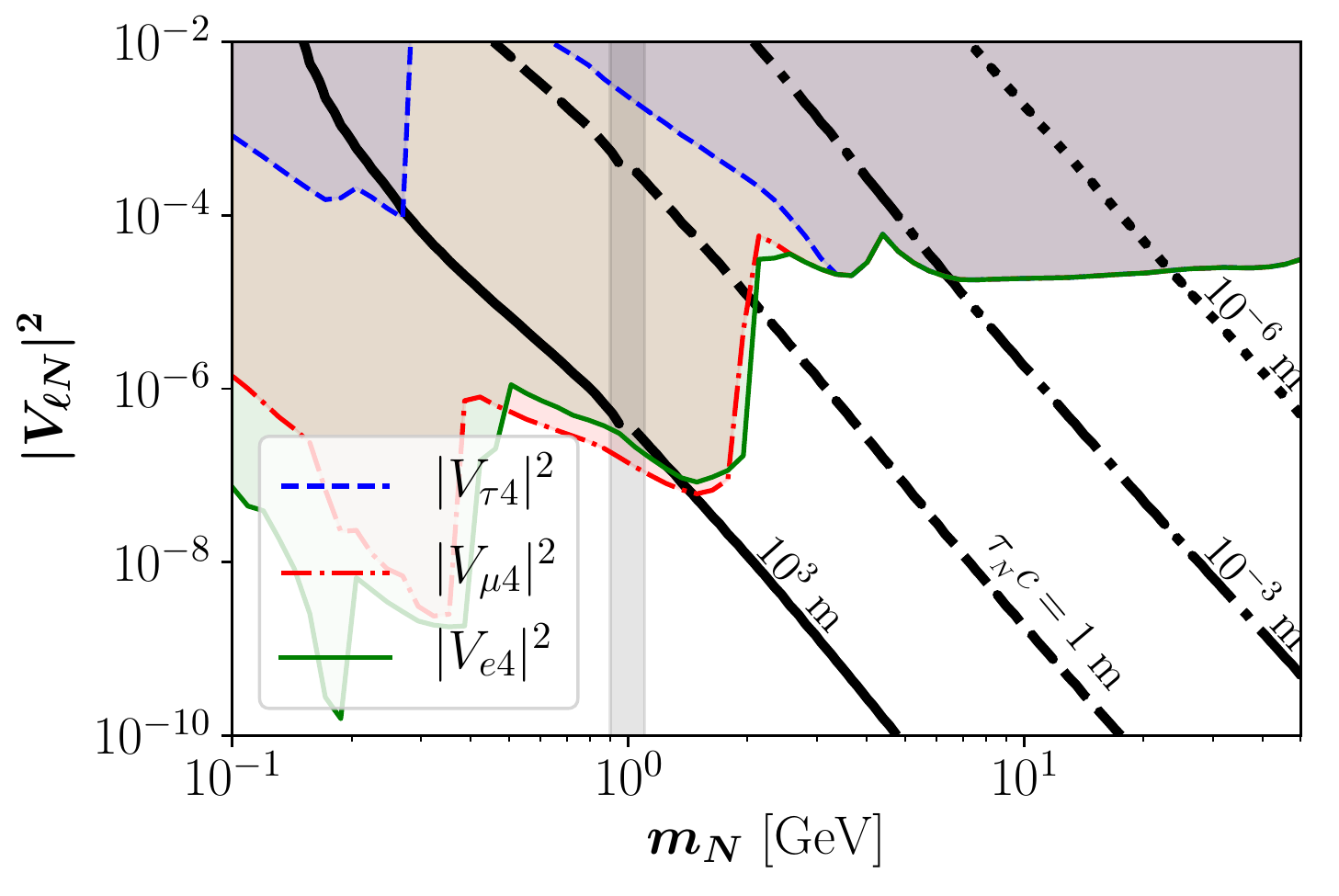}
\caption{Left: HNL total decay width $\Gamma_N$ as a function of its mass $m_N$ for $|V_{eN}|=|V_{\mu N}|=|V_{\tau N}|\equiv |V_{\ell N}|$. 
Right: contour lines for HNL decay length $\tau_{_N} c$. Shadowed areas are excluded by different experiments.
Gray vertical line separates the two regimes for semileptonic decays.
}\label{Decay_mN}
\end{center}
\end{figure}

We show in Fig.~\ref{Decay_mN} (left) how the total neutrino width $\Gamma_N$ depends on the HNL mass $m_N$.
This figure is done for a normalized situation of $|V_{eN}|=|V_{\mu N}|=|V_{\tau N}|$, although we have checked that the general behavior could be approximated as
\begin{equation}\label{eq:Nwidth}
\Gamma_N\propto G_F^2\, m_N^5 \sum_{\ell=e,\mu,\tau} \big| V_{\ell N} \big|^2\,,
\end{equation}
up to corrections due to threshold effects.
This total width translates to a decay length $\tau_{_N} c$ as shown in Fig.~\ref{Decay_mN} (right), again for degenerate $|V_{eN}|=|V_{\mu N}|=|V_{\tau N}|\equiv|V_{\ell N}|$ mixings.
Black lines in this figure show contours for $\tau_{_N} c=10^3, 1, 10^{-3}$ and $10^{-6}$~m, while shadowed areas are experimentally excluded\,\cite{Abada:2017jjx}.
The vertical gray line illustrates the transition\cite{Abada:2018sfh} between the semileptonic decay $N\to \ell M$ and $N\to \ell q\bar q'$.
Interestingly, we see that the region with decay lengths relevant for DV searches at the LHC, $\tau_{_N} c\in$~[1mm, 1m], lies in the few GeV region, where present experimental constraints on the mixings are weaker. 
In this region strong bounds come from DELPHI\,\cite{Abreu:1996pa}, precisely by looking for DV signatures from $Z\to \nu N$ decays.

\begin{figure}[t!]
\begin{center}
\includegraphics[width=0.49\textwidth]{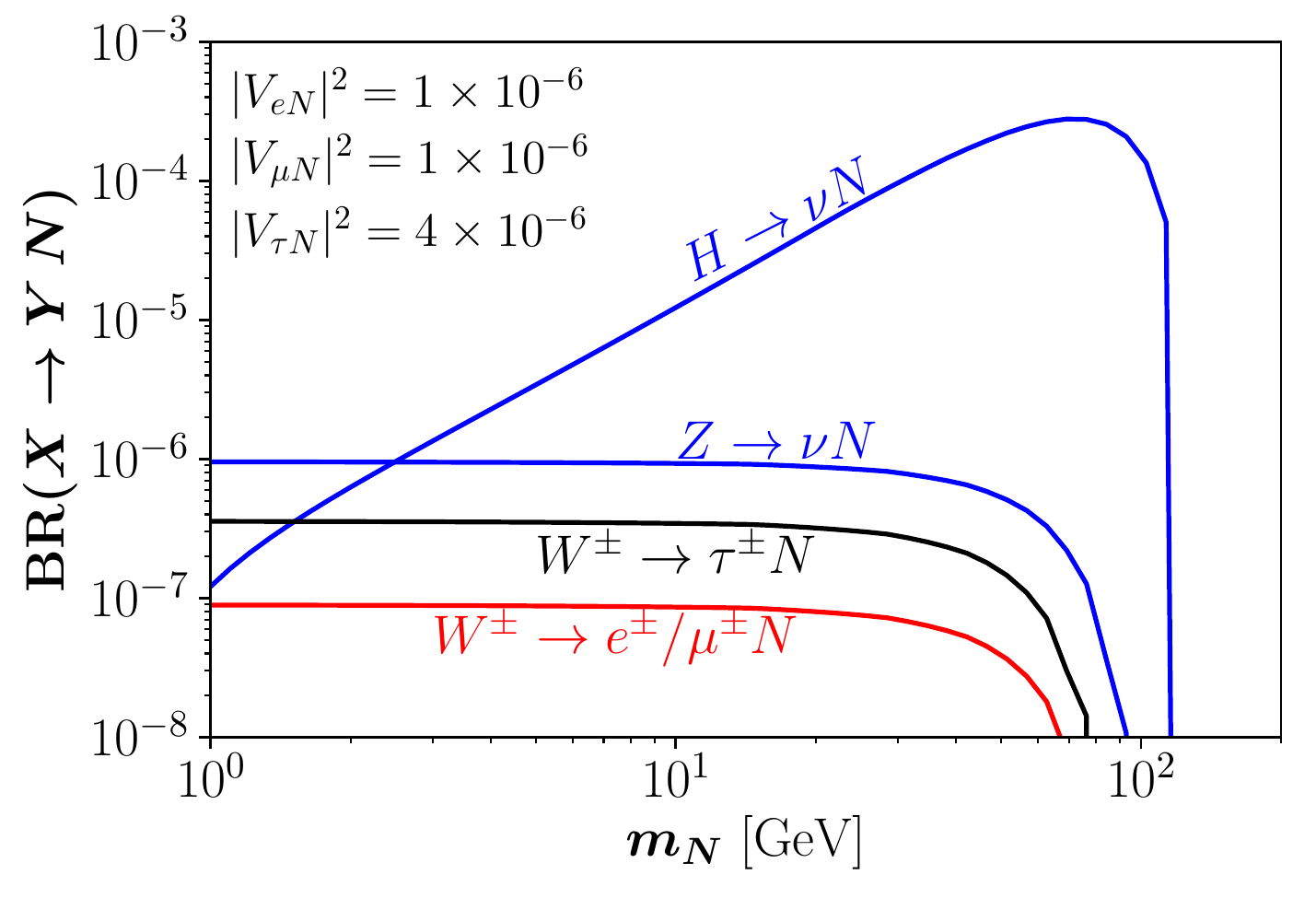}
\includegraphics[width=0.49\textwidth]{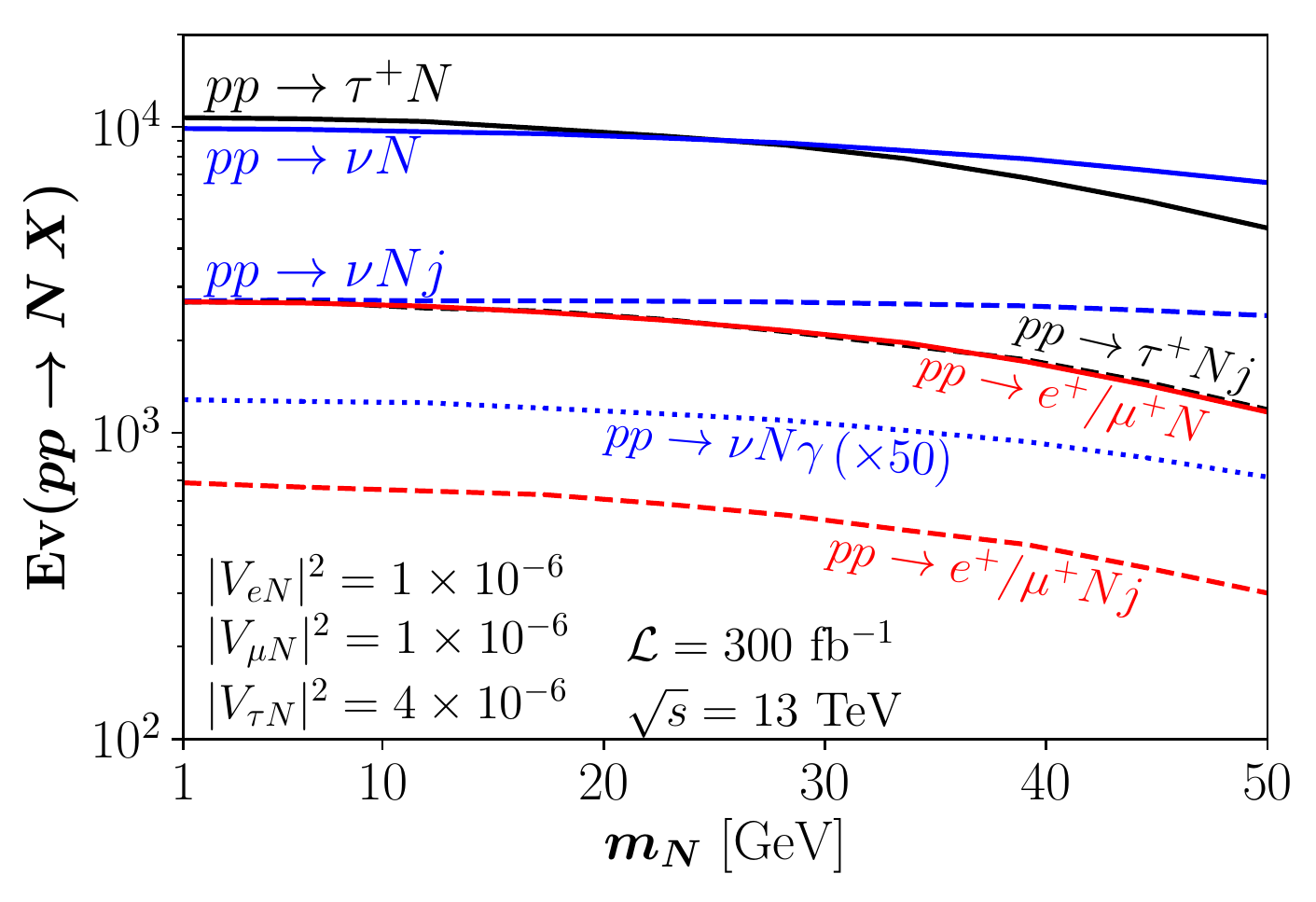}
\caption{Left: Branching ratios for $W\to\ell N$, $Z\to\nu N$ and $H\to\nu N$, where $\nu$ stands for the sum over the three light neutrinos. 
Right: HNL production at LHC in association with a charged lepton or light neutrinos. We also show the production with an additional jet or photon. 
}\label{Production-mN}
\end{center}
\end{figure}

HNL with masses of few GeV are mainly produced in $W$, $Z$ and $H$ decays. 
We display the branching ratio of each channel in Fig.~\ref{Production-mN} (left), from where we see that their behaviors with $m_N$ are different.
For the HNL masses of our interest, the decay of the $W$ and $Z$ gauge bosons are almost independent on $m_N$, since these decays can be well approximated as $W\to\ell \nu$ or $Z\to\nu\nu$ decays, followed by a $\nu$-$N$ mixing.
The case of the Higgs boson is clearly different and it can be understood as the chirality flip needed for the $H\to\nu N$ decay, which in this case is provided by the heaviest neutrino mass $m_N$.
Notice that the longitudinal components of the $W$ and $Z$ bosons also share this latter behavior, however their contribution to the BR is subleading due to the $m_N/m_W$ suppression.
All these features can be summarized as,
\begin{equation}
{\rm BR} (W^\pm\to \ell^\pm N)\propto \big| V_{\ell N}\big|^2\,,~
{\rm BR} (Z\to\nu N)\propto \sum_{\ell} \big| V_{\ell N}\big|^2\,,~
{\rm BR} (H \to\nu N)\propto \frac{m_N^2}{m_W^2}\,\sum_{\ell} \big| V_{\ell N}\big|^2\,,
\end{equation}
where we have summed over the three light neutrinos, as one cannot distinguish them. 
From these expressions we see that the HNL produced from $W$ decays, experimentally appealing due to the associated charged lepton, has a BR that  depends on each particular $V_{\ell N}$ mixing and therefore the sensitivity to these channels will be flavor dependent.
On the other hand, the production from $Z$ and $H$ decays depends on the sum of all the mixings and, consequently, they are flavor independent.
We will refer to these latter production channels as flavor blind production.
Notice that this combination of mixings is the same entering in the HNL total  width, see Eq.~(\ref{eq:Nwidth}), one of the most relevant parameters when studying DV signatures.

The LHC could help exploring this region of parameter space due to the huge amount of produced $W$, $Z$ and $H$ bosons.
We show in Fig.~\ref{Production-mN} (right) the number of HNL produced at $\mathcal L=300~{\rm fb}^{-1}$ of $pp$ collisions at $\sqrt{s}=13$~TeV.
We have used {\tt MadGraph5\_aMC@NLO}\,\cite{Alwall:2014hca}  to generate  the $pp\to\ell N$ and $pp\to\nu N$ processes and we have checked that they are indeed dominated by the production and decay of on-shell $W$ and $Z$ bosons, respectively. 
For the values of $m_N$ in which we will be interested in, we find that the production from $H$ decays is subdominant, although it may become relevant for heavier HNL. 
In this right panel we have chosen the same scenario as in the left one, and we can see that the production at $pp$ collisions follows the same pattern as the boson decays.
This means again that the relative importance of each $W$ channel depends on the relative size of each mixing, while the $Z$ channel is flavor blind and depends directly on the mass and total width of the HNL. 
As shown in this particular example, the flavor blind production could be of particular interest when the mixing to $\tau$ leptons is larger than those to electrons or muons, as the $\tau$ signatures are more difficult to handle in an hadronic collider. 

In Fig.~\ref{Production-mN} we show as well the events corresponding to the also flavor blind production channels\,\footnote{We apply for these channels minimum $p_T$ cuts of $p_T^j > 20$~GeV and $p_T^\gamma>10$~GeV.}  $pp\to\nu N j$ and $p p\to \nu N \gamma$.
These channels could be useful for the flavor blind production, as the initial state radiated jet or photon could be used as trigger for the interaction point in the $Z$ channel. 
We see from this figure that, despite they have lower cross-sections, they could be comparable to some charged channels if the HNL mixing to that flavor is small.

\section{Displaced Vertices at LHC from HNL decays}

We have seen that HNL with masses of few GeV and mixings allowed by present experimental constraints have decay lengths of the size of LHC detectors.
We have also learnt that at this mass scale the main production mechanism is from $W/Z/H$ boson decays, which are copiously produced at the LHC. 
Combining both, we can derive information on the LHC potential to probe this region of parameters space via DV searches.

\begin{figure}[t!]
\begin{center}
\includegraphics[width=0.49\textwidth]{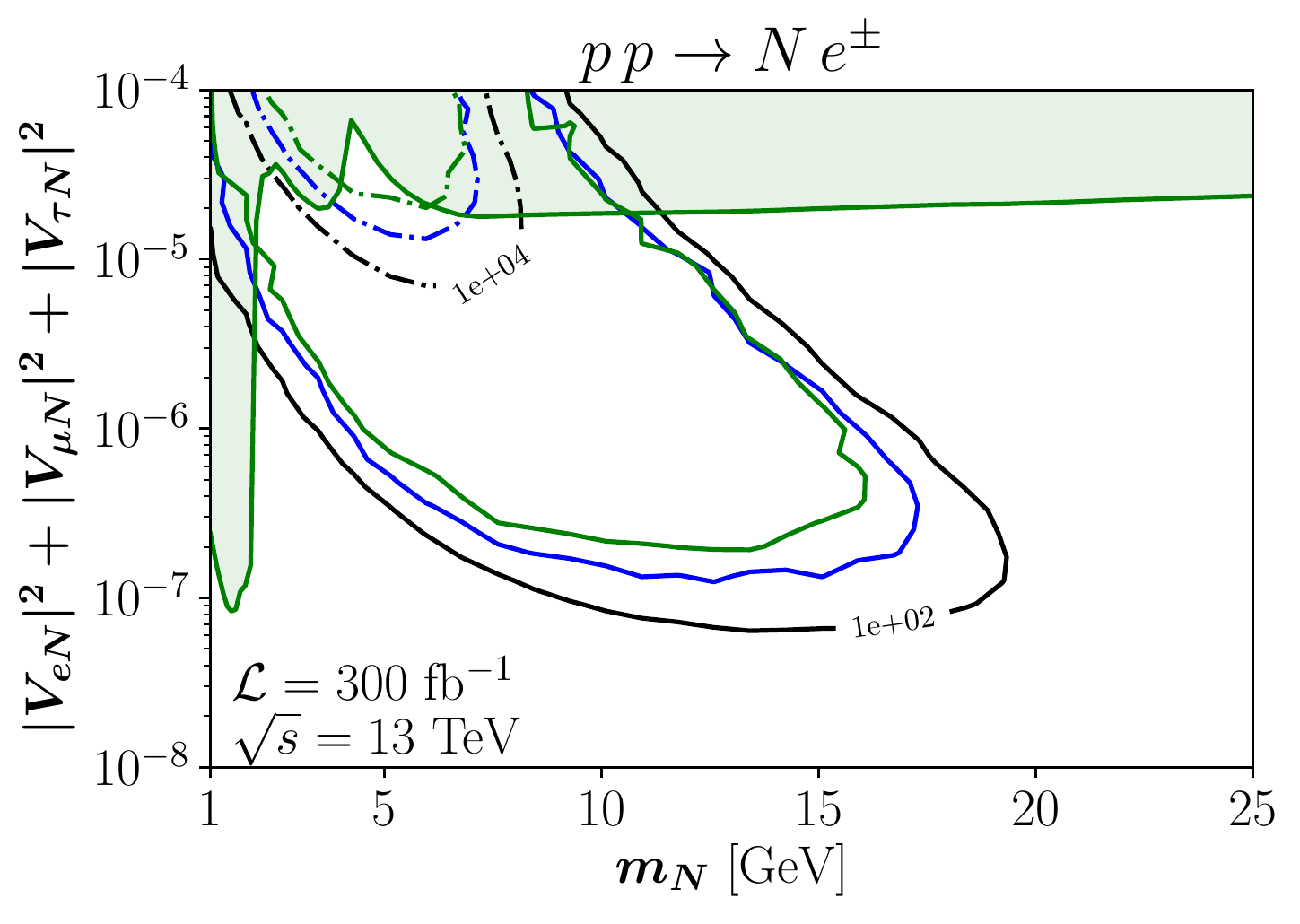}
\includegraphics[width=0.49\textwidth]{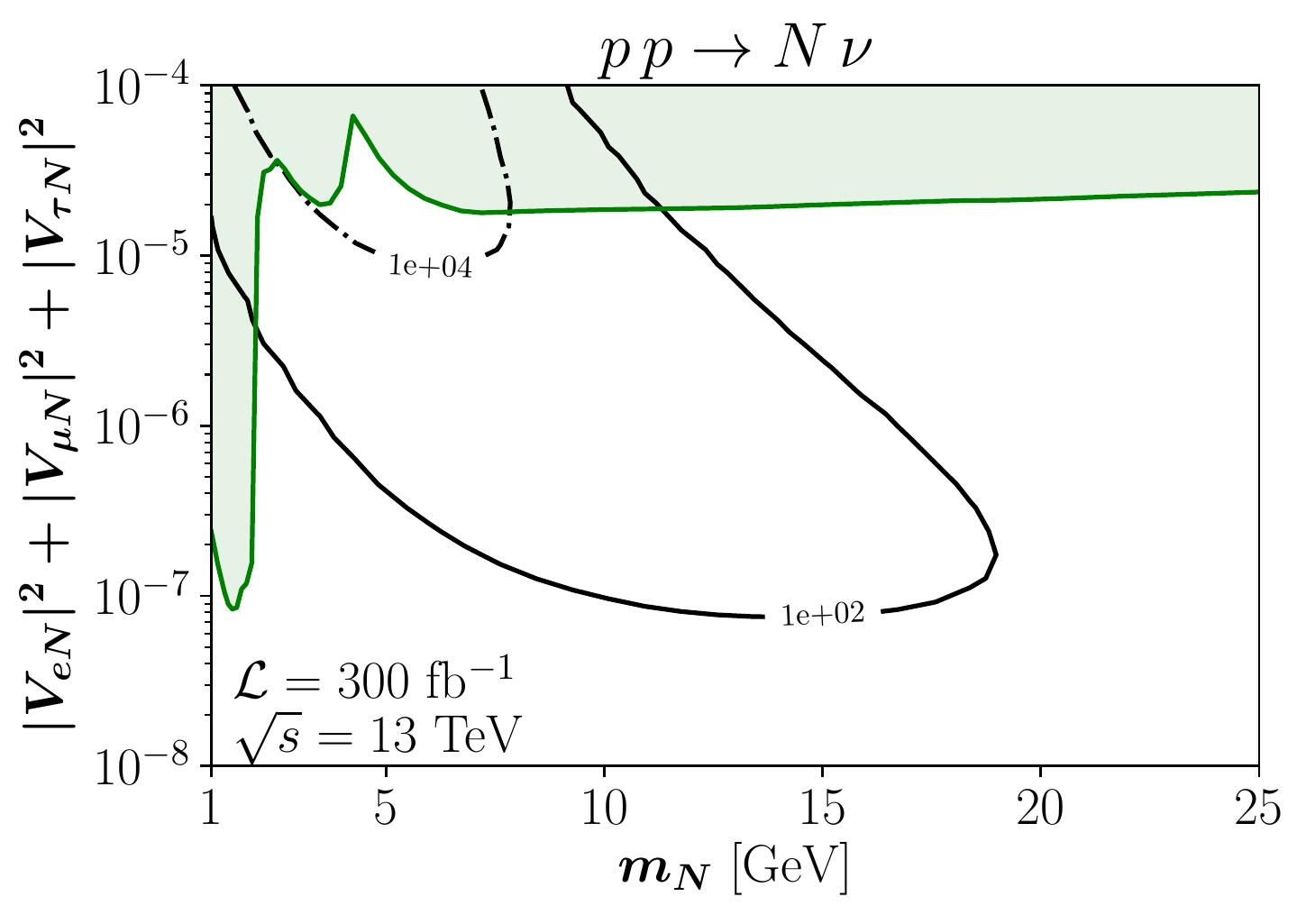}
\caption{Left: HNL production in association with a $e^\pm$, with $p_T^e>25$~GeV and $|\eta^e|<2.5$, for three scenarios: only one mixing $V_{eN}$ (black), two mixings $|V_{e N}|=|V_{\mu N}|$ (blue) and three mixings $|V_{e N}|=|V_{\mu N}|=|V_{\tau N}|$ (green). 
Right: HNL production in association with light neutrinos. Here the three scenarios give the same results. 
In both panels, solid  (dot-dashed) lines are contour lines for 100 (10000) events with DV at the LHC 13 TeV with $\mathcal L = 300~{\rm fb}^{-1}$.
Shadowed green area is excluded by experimental bounds.}\label{DVevents-mNVlN}
\end{center}
\end{figure}

In Fig.~\ref{DVevents-mNVlN} we show the number of DV events for $\mathcal L=300~{\rm fb}^{-1}$ coming from a HNL produced in the two kind of channels discussed above, i.e, $p p\to \ell N$ and $p p\to \nu N$.
We obtain the HNL total width by computing the main decay channels\,\cite{Abada:2017jjx,Bondarenko:2018ptm} and we implement it in {\tt MadGraph5\_aMC@NLO} to generate the DV topology. 
Thus,  Fig.~\ref{DVevents-mNVlN} corresponds to the total number of produced HNL that would decay to any channel at a second vertex separated from the interaction point. 
More precisely, we consider it as a DV if the displacement in the transverse plane is between 1~mm and 1~m from the primary vertex and less than 300~mm in the beam direction.

In the left panel we consider the HNL production in association with an electron or a positron.
This channel is very interesting since the charged lepton, if detected, could be used as a trigger for the primary vertex. 
We see that many events of this type could be produced for $\mathcal L=300~{\rm fb}^{-1}$, implying that the LHC could explore this mass range beyond present constraints, corresponding to the shadowed green area in these plots. 
Indeed, this production mechanism has been explored before\,\cite{Helo:2013esa,Izaguirre:2015pga} with the conclusion that the LHC could probe mixings up to $|V_{\ell N}|\sim10^{-7}$, with $\ell=e, \mu$, after collecting $\mathcal L=300~{\rm fb}^{-1}$ of data.

Nevertheless, it is important to notice that these works assumed that the HNL mixes only to one flavor, which is not the general situation in most of the BSM models. 
In order to show the effect of deviating from this simplified hypothesis, we show in both panels of Fig.~\ref{DVevents-mNVlN}  the results for three different benchmark scenarios:
\begin{enumerate}
\item Black lines: Mixing to one flavor, $V_{e N}$, with $V_{\mu N}=V_{\tau N}=0$.
\item Blue lines: Mixing to two flavors, $\big|V_{e N}\big|=\big|V_{\mu N}\big|$, with $V_{\tau N} =0$.
\item Green lines: Democratic mixing to three flavors $\big|V_{e N}\big|=\big|V_{\mu N}\big|=\big|V_{\tau N}\big|$.
\end{enumerate}
From the left panel we learn that the number of events, and therefore the sensitivity, is different in the three explored scenarios, the most optimistic numbers being those of the simplified scenario with only one non-zero mixing. 
The differences come from the fact that the total width, defining the area where DV may occur, depends on the sum of all mixings, as already explained in Eq.~(\ref{eq:Nwidth}), while the production rate of $p p\to e N$ is only sensitive to $|V_{e N}|$. 
Therefore, when considering mixings also to other flavors, the relative importance of $|V_{eN}|$ decreases and so does the sensitivity via the $p p\to eN$ channel.
Similar conclusions apply to searches for other charged leptons.

We have performed  the same exercise for the $p p\to \nu N$ production channel and we have observed that the three considered scenarios lead to the same number of  DV events, shown as a single color in the right panel  of Fig.~\ref{DVevents-mNVlN}.
The reason is that in this case, as we explained before, both the total width and the production rate depend on the same combination $|V_{eN}|^2+ |V_{\mu N}|^2+ |V_{\tau N}|^2$.
Therefore, this flavor blind production mechanism is closely related to DV searches, as it was already pointed out when exploring HNL via $H$ decays\,\cite{Gago:2015vma}.
Of course, a flavor dependence will enter if the HNL decays to charged leptons, nevertheless it will be milder than in the $W$ production channel,  where the flavor dependence enters  in both production and decay of the HNL.

In comparison with the $p p\to \ell N$ channel, the flavor blind production has the disadvantage of not having anything to trigger on in the primary vertex. 
A possible improvement to this situation could be to consider initial state radiation of jets or photons  to trigger the primary vertex where the HNL was produced. 
As shown in Fig.~\ref{Production-mN}, the cross sections  for these processes are smaller but still comparable to other channels, depending on the mixings. 
Nevertheless, a more dedicated study is needed, so we refer to the main paper\,\cite{Abada:2018sfh} for this discussion.

\section{Conclusions}

Heavy neutral leptons are introduced by many well-motivated models in order to solve  open problems in the SM, thus it is mandatory to look for all possible phenomenological consequences in order to test these theories.
We have seen that HNL with masses of few GeV and with mixings allowed by present experimental constraints could lead to displaced vertex signals at the LHC, which are clear signatures due to the low SM backgrounds. 
We have discussed the different production mechanisms, emphasizing the impact of the flavor structure of the HNL and the advantages of considering flavor blind production mechanisms in DV searches.  
In the end, we conclude that it is important to look for all possible channels, as they are complementary and necessary to explore the full parameter space.

\section*{Acknowledgments}

X.M. thanks the Moriond organizing committee for its financial support to attend the conference.
This work is supported  by the European Union Horizon 2020
research and innovation programme under the Marie Sk{\l}odowska-Curie: RISE
InvisiblesPlus (grant agreement No 690575)  and 
the ITN Elusives (grant agreement No 674896).


\section*{References}

\begin{thebibliography}{99}

\bibitem{Atre:2009rg}
  A.~Atre {\it et al.}
 {\it   JHEP {\bf 0905} (2009) 030}
   
\bibitem{Abada:2017jjx}
  A.~Abada {\it et al.}
{\it   JHEP {\bf 1802} (2018) 169}

\bibitem{Nemevsek:2011hz}
  M.~Nemevsek {\it et al.}
  {\it Phys.\ Rev.\ D {\bf 83} (2011) 115014}
  
\bibitem{Helo:2013esa}
  J.~C.~Helo, {\it et al.}
{\it  Phys.\ Rev.\ D {\bf 89} (2014) 073005}

\bibitem{Izaguirre:2015pga}
  E.~Izaguirre {\it et al.}
 {\it    Phys.\ Rev.\ D {\bf 91} (2015) no.9,  093010}
   
\bibitem{Dube:2017jgo}
  S.~Dube {\it et al.}
 {\it  Phys.\ Rev.\ D {\bf 96} (2017) no.5,  055031}
  
\bibitem{Cottin:2018kmq}
  G.~Cottin {\it et al.}
  {\it   Phys.\ Rev.\ D {\bf 97} (2018) no.5,  055025}
 
\bibitem{Cottin:2018nms}
  G.~Cottin {\it et al.}
  arXiv:1806.05191 [hep-ph].

\bibitem{Dib:2018iyr}
  C.~O.~Dib {\it et al.}
  {\it Phys.\ Rev.\ D {\bf 97} (2018) no.3,  035022}
 
\bibitem{Nemevsek:2018bbt}
  M.~Nemevsek  {\it et al.}
  {\it Phys.\ Rev.\ D {\bf 97} (2018) no.11,  115018}

\bibitem{Maiezza:2015lza}
  A.~Maiezza {\it et al.}
  {\it Phys.\ Rev.\ Lett.\  {\bf 115} (2015) 081802}
   
\bibitem{Gago:2015vma}
  A.~M.~Gago {\it et al.}
{\it  Eur.\ Phys.\ J.\ C {\bf 75} (2015) no.10,  470}

\bibitem{Accomando:2016rpc}
  E.~Accomando  {\it et al.}
  {\it JHEP {\bf 1704} (2017) 081 }

\bibitem{Nemevsek:2016enw}
  M.~Nemevsek  {\it et al.}
  {\it JHEP {\bf 1704} (2017) 114}

\bibitem{Caputo:2017pit}
  A.~Caputo {\it et al.}
  {\it JHEP {\bf 1706} (2017) 112}

\bibitem{Deppisch:2018eth}
  F.~F.~Deppisch  {\it et al.}
  arXiv:1804.04075 [hep-ph].


\bibitem{Antusch:2017hhu}
  S.~Antusch {\it et al.}
{\it  Phys.\ Lett.\ B {\bf 774} (2017) 114}
  
\bibitem{Caputo:2016ojx}
  A.~Caputo  {\it et al.}
  {\it Eur.\ Phys.\ J.\ C {\bf 77} (2017) no.4,  258}

\bibitem{Kling:2018wct}
  F.~Kling {\it et al.}
{\it   Phys.\ Rev.\ D {\bf 97} (2018) no.9,  095016 }

  
\bibitem{Helo:2018qej}
  J.~C.~Helo {\it et al.}
{\it  JHEP {\bf 1807} (2018) 056 }

\bibitem{Jana:2018rdf}
  S.~Jana {\it et al.}
  arXiv:1804.06828 [hep-ph].
  
\bibitem{Curtin:2018mvb}
  D.~Curtin {\it et al.}
  arXiv:1806.07396 [hep-ph].
   
\bibitem{Abada:2018sfh}
  A.~Abada, N.~Bernal, M.~Losada and X.~Marcano,
  arXiv:1807.10024 [hep-ph].
  
\bibitem{Bondarenko:2018ptm}
  K.~Bondarenko {\it et al.}
  arXiv:1805.08567 [hep-ph].
  
\bibitem{Abreu:1996pa}
  P.~Abreu {\it et al.} [DELPHI Collaboration],
 {\it Z.\ Phys.\ C {\bf 74} (1997) 57
   Erratum: [Z.\ Phys.\ C {\bf 75} (1997) 580]}
  
\bibitem{Alwall:2014hca}
  J.~Alwall {\it et al.}
{\it  JHEP {\bf 1407} (2014) 079}
  
\end{thebibliography}
\end{document}